\documentclass[twocolumn]{aastex63}

\usepackage[T1]{fontenc}
\usepackage{bm}        
\usepackage{amssymb, amsmath}   
\usepackage{cancel}
\usepackage{color}

\DeclareMathAlphabet{\mathsfit}{\encodingdefault}{\sfdefault}{m}{sl}
\SetMathAlphabet{\mathsfit}{bold}{\encodingdefault}{\sfdefault}{bx}{sl}
\newcommand{\vect}[1]{\bm{#1}}

\DeclareMathOperator{\sech}{sech}
\shorttitle{X-ray Polarization from Reconnection}
\shortauthors{Zhang et al.}

\begin{document}

\title{First-Principle Prediction of X-ray Polarization from Magnetic Reconnection in High-Frequency BL Lacs}

\correspondingauthor{Haocheng Zhang}
\email{astrophyszhc@hotmail.com}

\author[0000-0001-9826-1759]{Haocheng Zhang}
\affiliation{Department of Physics and Astronomy \\
Purdue University \\
West Lafayette, IN 47907, USA}
\affiliation{New Mexico Consortium \\
Los Alamos, NM 87544, USA}

\author[0000-0001-5278-8029]{Xiaocan Li}
\affiliation{Dartmouth College\\
Hanover, NH 03750, USA}

\author[0000-0003-1503-2446]{Dimitrios Giannios}
\affiliation{Department of Physics and Astronomy \\
Purdue University \\
West Lafayette, IN 47907, USA}

\author[0000-0003-4315-3755]{Fan Guo}
\affiliation{Theoretical Division \\
Los Alamos National Lab \\
Los Alamos, NM 87545, USA}
\affiliation{New Mexico Consortium \\
Los Alamos, NM 87544, USA}

\begin{abstract}
Relativistic magnetic reconnection is a potential particle acceleration mechanism for high-frequency BL Lacs (HBLs). The {\it Imaging X-ray Polarimetry Explorer} ({\it IXPE}) scheduled to launch in 2021 has the capability to probe the magnetic field evolution in HBLs, examining the magnetic reconnection scenario for the HBL flares. In this paper, we make the first attempt to self-consistently predict HBL X-ray polarization signatures arising from relativistic magnetic reconnection via combined particle-in-cell (PIC) and polarized radiation transfer simulations. We find that although the intrinsic optical and X-ray polarization degrees are similar on average, the X-ray polarization is much more variable in both polarization degree and angle (PD and PA). Given the sensitivity of the {\it IXPE}, it may obtain one to a few polarization data points for one flaring event of nearby bright HBLs Mrk~421 and 501. However, it may not fully resolve the highly variable X-ray polarization. Due to the temporal depolarization, where the integration of photons with variable polarization states over a finite period of time can lower the detected PD, the measured X-ray PD can be considerably lower than the optical counterpart or even undetectable. The lower X-ray PD than the optical thus can be a characteristic signature of relativistic magnetic reconnection. For very bright flares where the X-ray polarization is well resolved, relativistic magnetic reconnection predicts smooth X-ray PA swings, which originate from large plasmoid mergers in the reconnection region.
\end{abstract}



\section{Introduction} \label{sec:intro}

HBLs, such as Mrk~421 and 501, are among the most powerful particle accelerators in the Universe. They exhibit variable nonthermal-dominated emission up to TeV $\gamma$-rays, with the flaring time scale as short as a few minutes in the TeV band, indicating extreme particle acceleration in very localized regions \citep{Albert2007a,Albert2007b}. Their emission originates from relativistic jets pointing very close to our line of sight. Their spectral energy distribution (SED) has two components: the low-energy component due to synchrotron by highly relativistic electrons peaks at soft X-rays, which gives them the name ``high-frequency''; the high-energy component extends from X-ray to TeV $\gamma$-ray, which is often considered as Compton scattering by the same electrons that make the synchrotron component \citep{Padovani1995}.

Relativistic magnetic reconnection is a candidate particle acceleration mechanism for HBL emission. During this plasma physics process, oppositely directed magnetic field lines break and rejoin, dissipating a large amount of magnetic energy. Recent simulations have suggested that reconnection can efficiently accelerate particles into power-law distributions in a magnetized environment \citep[][and see \citet{Guo2020a} for a recent review]{Guo2014,Guo2016,Sironi2014,Werner2016,Li2018,Li2019}. Additionally, radiation from relativistic outflows in the reconnection region may experience additional relativistic boosting, making it a very attractive scenario for the extreme TeV variability \citep{Giannios2009,Sironi2016,Christie2020}. Nevertheless, so far we lack distinct observable signatures from reconnection that can pinpoint its presence in the HBL emission region.

The scheduled launch of the {\it IXPE\footnote{\url{https://ixpe.msfc.nasa.gov/}}} will open up a unique window to study HBLs via X-ray polarimetry. Recent numerical simulations of reconnection have shown characteristic optical polarization patterns \citep{Zhang2018,Zhang2020,Hosking2020}. If similar patterns exist in the X-ray polarization, the {\it IXPE} can unveil the magnetic field structure and evolution during HBL flares and identify potential reconnection processes. However, the HBL X-ray emission is highly variable. If its polarization is also variable, the integration of photons with different polarization states over time can diminish the detected PD. This ``temporal depolarization'' must be properly considered in predicting X-ray polarization.

In this paper, we make the first attempt to self-consistently predict the time-dependent X-ray polarization from relativistic magnetic reconnection in HBLs. We use combined PIC and polarized radiation transfer simulations to model the magnetic field and particle evolution under first principles and to include temporal depolarization via ray-tracing. We aim to identify characteristic X-ray polarization patterns from reconnection by comparison with the optical counterpart, which already has rich observational data \citep{Hovatta2016,Fraija2017,Aleksic2015}. Section \ref{sec:tempdepol} describes the temporal depolarization effect, Section \ref{sec:simulation} presents our simulation setup and results, and Section \ref{sec:discussion} discusses implications for optical and X-ray polarimetry.

\section{Temporal Depolarization \label{sec:tempdepol}}

\begin{figure}
\centering
\includegraphics[width=0.493\linewidth]{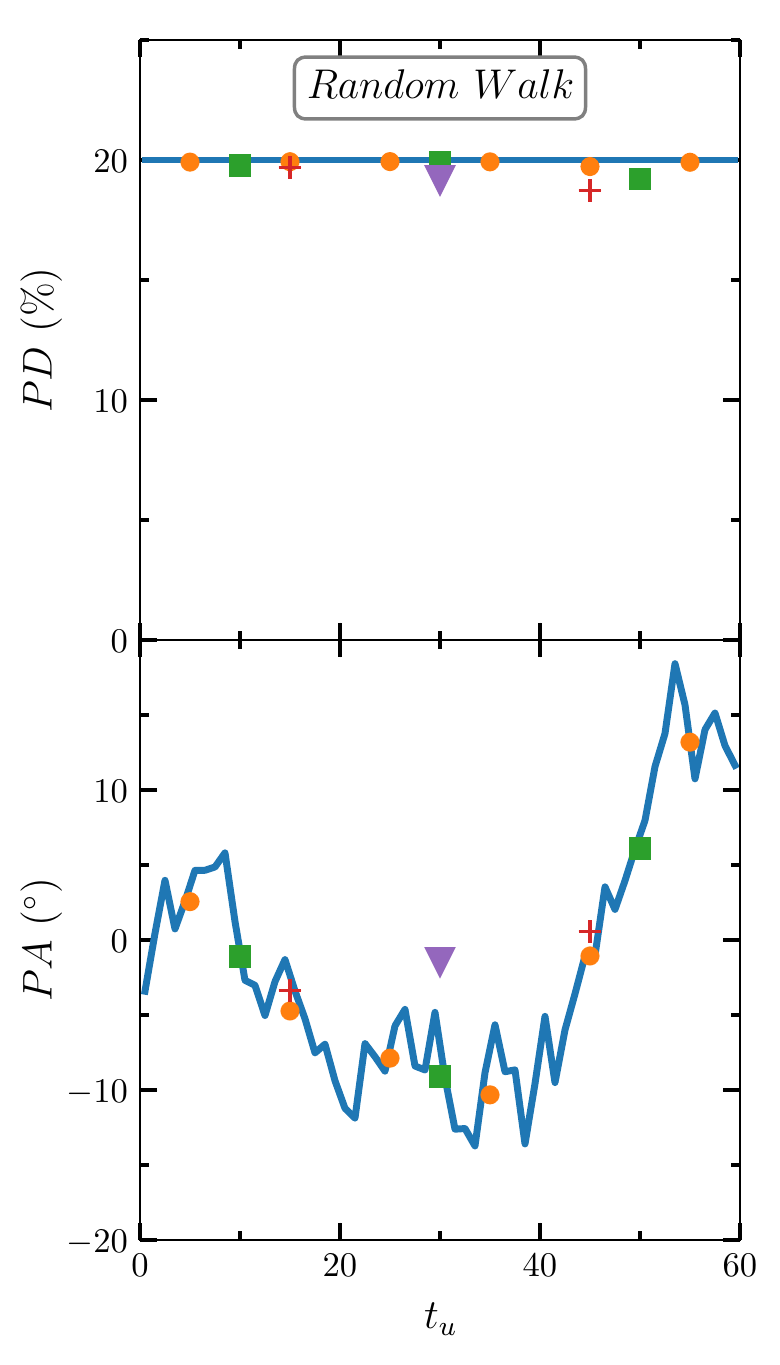}
\includegraphics[width=0.493\linewidth]{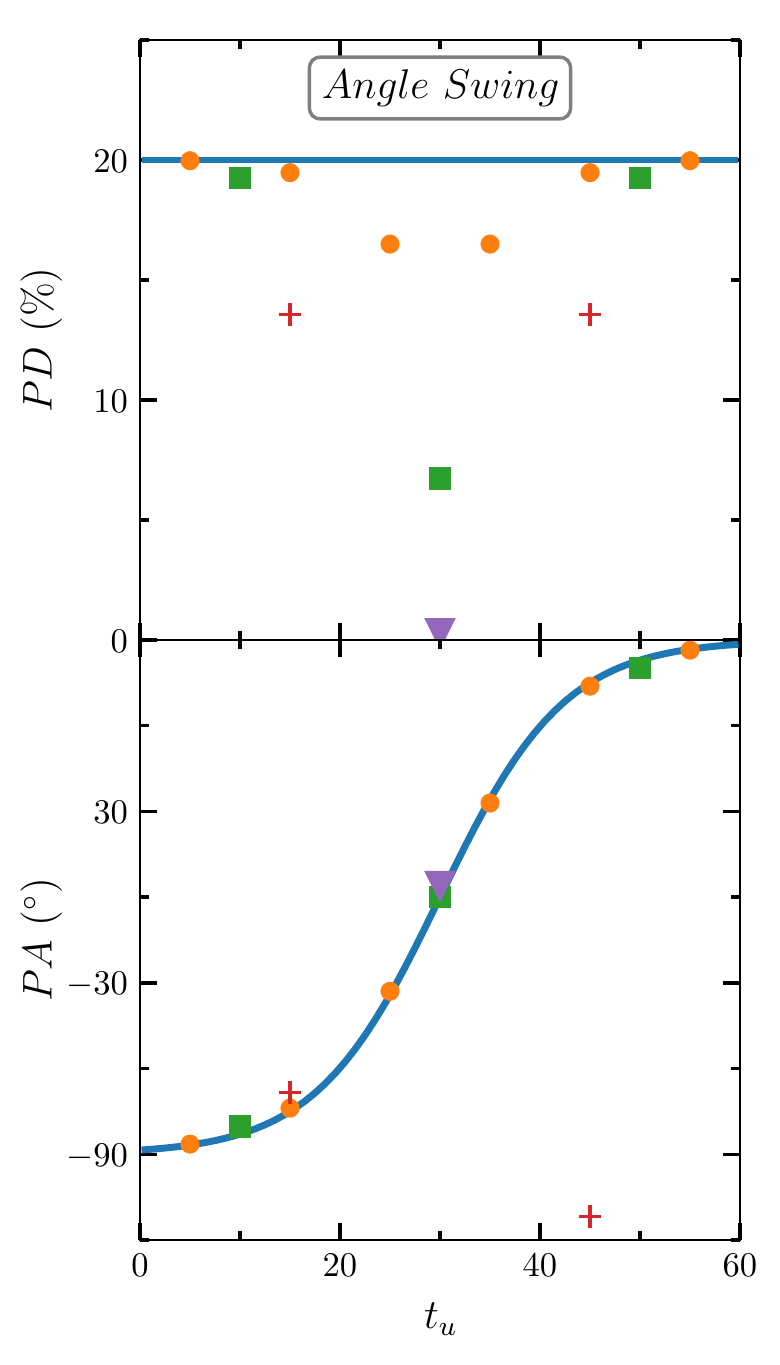}
\caption{Two examples to illustrate the temporal depolarization. The left panels show randomly fluctuating PA, while the right panels show a PA swing event (intrinsic evolution shown in blue curves). The top and bottom panels show the PDs and PAs, respectively. The orange circles, green squares, red crosses, and the purple triangles show the detected polarization with different temporal resolution (six, three, two, and one point, respectively).}
\label{fig:tempdepol}
\end{figure}

Temporal depolarization can considerably reduce the detected PD up to $100\%$ with respect to the intrinsic value. This comes from the integration of photons with different polarization states from a variable source over time. Observations have shown that the optical polarization of blazars, including HBLs, can be variable, especially during flares \citep{Blinov2016,Fraija2017}. This implies magnetic field evolution in the emission region. Therefore, the X-ray polarization, also originating from synchrotron emission, can be variable. As X-ray photons carry much more energy than the optical, even though the X-ray flux is higher than the optical for HBLs, the number of X-ray photons per unit time received by the telescope is much less than the optical. As a result, the {\it IXPE} is unlikely to obtain so high temporal resolution as the optical polarimeters. The unresolved X-ray polarization variations then may diminish the detected polarization via temporal depolarization.

We suggest that the temporal depolarization depends on the amplitude of the unresolved PA variation. Here we illustrate this dependence with two simple examples (Figure \ref{fig:tempdepol}). Both events have constant flux and $\rm{PD}=20\%$ that last 60 time units $t_u$. The first one has randomly fluctuating PA with a maximal change of $10^{\circ}$ every $t_u$. The second one undergoes a full $180^{\circ}$ PA swing from $-90^{\circ}$ to $90^{\circ}$. If the two events can only be resolved by a few data points, then the second event generally experiences larger unresolved PA variation per detection. Thus the second event has lower detected PD than the intrinsic. In particular, if the whole swing is only resolved as one data point, then the detected PD drops to zero; with two points, the PA swing cannot be resolved (Figure \ref{fig:tempdepol}). For both events, the higher the temporal resolution is, the closer the detected PD and PA can approach the intrinsic value.

The reason lies in the incoherent addition of emission with different polarization states. The detected PD and PA in a period of $t$ can be calculated by the Stokes parameters (we only consider the linear PD and PA given the synchrotron emission),
\begin{equation}
\begin{aligned}
\rm{PD}&= \frac{\sqrt{(\int_0^t Q(t')dt')^2+\int_0^t U(t')dt')^2}}{\int_0^tI(t')dt'} \\
\rm{PA}&= \tan^{-1}\frac{\int_0^t U(t')dt'}{\int_0^t Q(t')dt'}
\end{aligned}~~,
\end{equation}
where $I(t')$, $Q(t')$, and $U(t')$ are the intrinsic Stokes parameters at $t'$. Since the intrinsic flux and PD are constant for both events, we take $I=1$ and $\sqrt{Q^2+U^2}=0.2$ for every $t_u$. If the PA slightly fluctuates around $\rm{PA}=0$ as in the first event, at every $t_u$ $Q\lesssim 0.2$ and $U\sim 0$. By integrating over $t$, the total $U$ averages to zero, while the total $Q\sim 0.2t$. Thus the total PD is still nearly $20\%$ even if the entire event is unresolved. In the second event, the Stokes $Q$ moves from -0.2 to 0.2 then back to -0.2, while the Stokes $U$ changes from 0 to 0.2 then to -0.2 and back to 0. If the temporal resolution is not high enough, especially if the whole event is unresolved, the integration of $Q$ and $U$ are both zero, thus the detected PD is zero. With the high temporal resolution (orange circles in Figure \ref{fig:tempdepol}), the unresolved PA variation is small per detection, thus the integration does not significantly divert detected PD and PA from intrinsic values. Additionally, since the PA has $180^{\circ}$ ambiguity, observations generally consider the PA difference between two consecutive polarization data points to be less than $90^{\circ}$. Therefore, when the time resolution is low (the red cross in Figure \ref{fig:tempdepol} lower right), the PA swing is unresolved.

\section{Optical and X-ray Polarization from Reconnection \label{sec:simulation}}

Due to the drastically different radiative cooling of the optical and X-ray emission in HBLs, we expect distinct polarization from these two bands. In this section, we use PIC simulation to self-consistently study the highly dynamical evolution of the magnetic field and nonthermal particles during reconnection \citep{Zhang2018,Zhang2020}, and use ray-tracing polarized radiation transfer to simulate the HBL radiation and polarization signatures \citep{Zhang2015,Zhang2017}, including all time-dependent effects such as the temporal depolarization.

\subsection{Simulation Setup \label{sec:setup}}

We assume a preexisting current sheet in the HBL flaring region. Such structures may exist, for instance, under the striped jet model \citep{Giannios2019,Zhang2021}. Many HBLs have higher luminosity in the synchrotron spectral component than the high-energy component, indicating that the cooling by Compton scattering is subdominant \citep{Finke2013}. Since the keV X-ray emission from Mrk~421 and 501 that can be detected by the {\it IXPE} is synchrotron, our combined simulations only consider the synchrotron radiation and cooling for simplicity.

The PIC simulation setup is very similar to previous studies \citep{Zhang2018,Zhang2020,Kilian2020,Liu2020}. Here we describe some of the key parameters. We perform the 2D PIC simulation in the $x$-$z$ plane using the \texttt{VPIC} code \citep{Bowers2008}. The simulation assumes an electron-ion plasma with realistic mass ratio $m_i/m_e=1836$. The initial particle distributions are Maxwell–J\"uttner distributions with uniform density $n_0$ and temperature $T_e=T_i=400 m_e c^2$. This value is generally consistent with the typical low-energy cutoff of the particle spectral distributions based on blazar spectral fitting models \citep{Chen2011,Ahnen2018}. The upstream thermal electron inertial length is then $d_e=\sqrt{1+3T_e/(2m_ec^2)}d_{e0}\sim 24.5d_{e0}$. Reconnection starts from a magnetically-dominated force-free current sheet, $\vect{B}=B_0\tanh(z/\lambda)\hat{x}+B_0\sqrt{\sech^2(z/\lambda)+B_g^2/B_0^2}\hat{y}$, where $B_g=0.2B_0$ is the strength of the guide field, which is the component perpendicular to the anti-parallel components $B_0$. The half-thickness of the current sheet is $\lambda=0.6\sqrt{\sigma_e}d_{e0}$, where $d_{e0}=c/\omega_{pe0}$ is the nonrelativistic electron inertial length, $\omega_{pe0}=\sqrt{4\pi n_ee^2/m_e}$ is the nonrelativistic electron plasma frequency, and $\sigma_e=B_0^2/(4\pi n_em_ec^2)$ is the cold electron magnetization parameter \citep{Sironi2014,Guo2014}. We choose $\sigma_e=6.4\times 10^5$, corresponding to a total magnetization of $\sigma_0\approx(m_e/m_i)\sigma_e\approx 350$. We choose this value so that the upstream electron temperature is much lower than the cold electron magnetization parameter, while the electron spectrum can grow to the typical high-energy spectral cutoff ($\gamma_{max}\sim 10^5$ based on spectral fitting models) within a reasonably short period of simulation time. We note that the initial magnetization factor is in the upstream plasma, but the emission region of the reconnection is in the downstream plasma, where the magnetization factor is approximately one. The simulation box size is $2L\times L$ in the $x{\text -}z$ plane, where $L=16000 d_{e0}\sim 653d_e$. The $x$-axis has periodic boundaries for both fields and particles, while the $z$-axis has conductive boundary for fields but reflects particles. The simulation grid size is $4096\times2048$, with $100$ electron-ion pairs in each cell. The cell size $\Delta x=\Delta z\sim 0.32d_e$ can resolve the upstream electron inertial length. We mimic the synchrotron cooling effect by implementing a radiation reaction force. The strength is set so that the cooling spectral break happens at $\sim 1~\rm{keV}$, consistent with Mrk~421 and 501 observations \citep{Albert2007a,Albert2007b}.

We use the \texttt{3DPol} code to perform polarized radiation transfer simulations \citep{Zhang2014}. It evaluates the polarized synchrotron emission from each cell based on the magnetic field and particle distributions from PIC, then traces the emission to the plane of sky. We fix our line of sight along the $y$-axis. We normalize $B_0$ to $0.1~\rm{G}$ and give the simulation domain a bulk Lorentz factor $\Gamma=10$ in the $z$-direction. These parameters are chosen to be consistent with typical spectral fitting parameters for Mrk~421 and 501 \citep{Blazejowski2005}. The light crossing time of the simulation box in $x$-direction is $\tau_{lc}=32000t_0$, where $t_0=d_{e0}/c$. We output the particle and magnetic field from PIC every $250t_0 \sim 0.0078\tau_{lc}$ to closely follow the reconnection evolution. We do not include local Lorentz factor in the radiation transfer because the plasmoid motion is generally non-relativistic, due to the periodic boundary condition in PIC.

\begin{figure}
\centering
\includegraphics[width=0.99\linewidth]{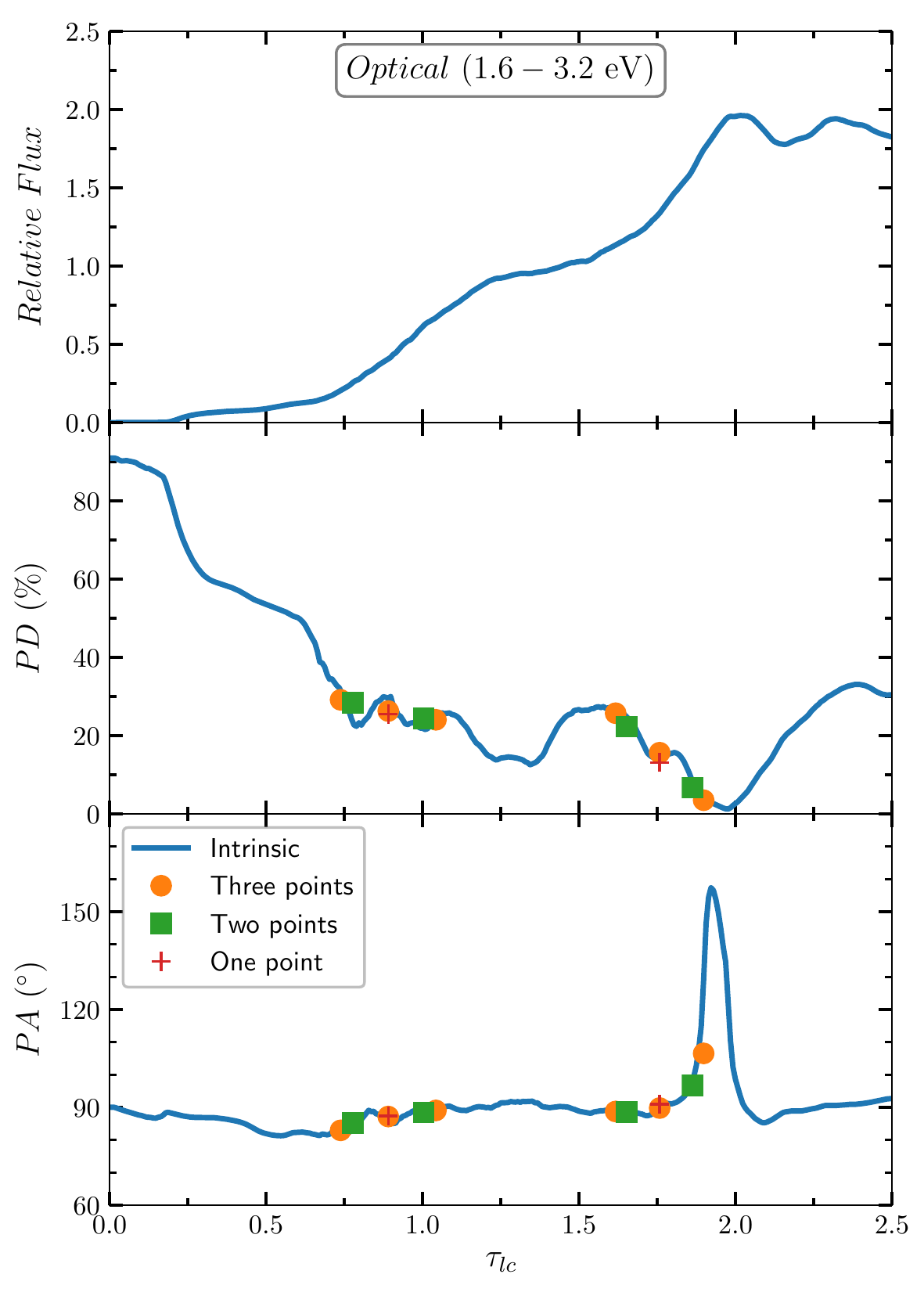}
\caption{From top to bottom are the light curve, PD and PA for the optical band from our combined simulation. Points represent the detected polarization with various temporal resolution during the two X-ray flares.}
\label{fig:optical}
\end{figure}

\begin{figure}
\centering
\includegraphics[width=0.99\linewidth]{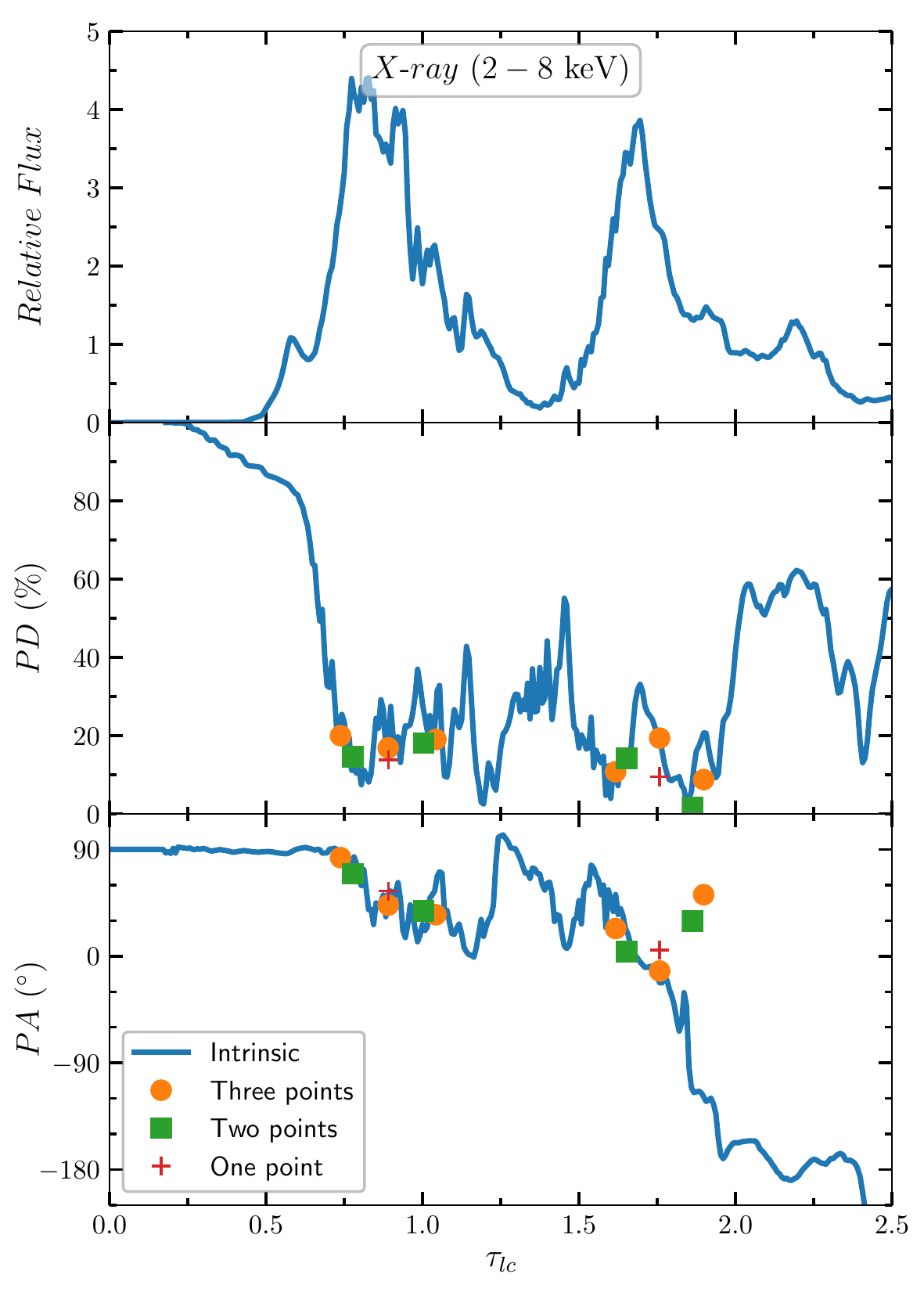}
\caption{Same as Figure \ref{fig:optical} but for the {\it IXPE} X-ray band.} 
\label{fig:xray}
\end{figure}

\subsection{Results \label{sec:results}}

\begin{figure*}
\centering
\includegraphics[width=0.99\textwidth]{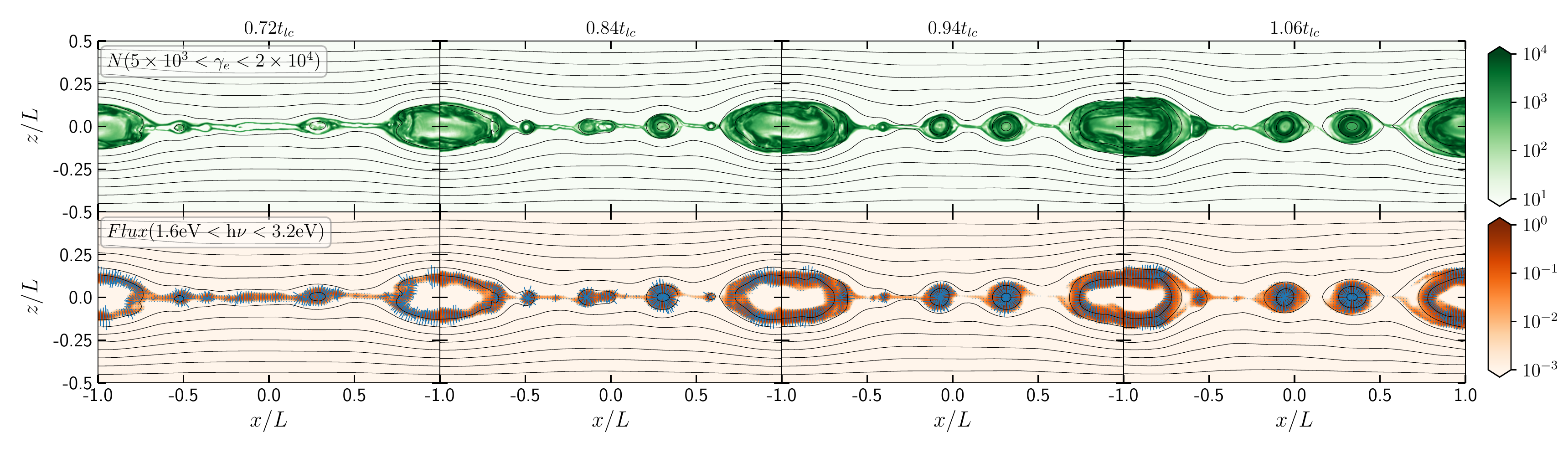}
\caption{From left to right are four snapshots of the particle spatial distribution (upper row) and polarized emission map (lower row) for the optical band during the first X-ray flare. The length and direction of the blue segments in the lower row represent the local polarized flux and angle, respectively.}
\label{fig:opticalsim}
\end{figure*}

\begin{figure*}
\centering
\includegraphics[width=0.99\textwidth]{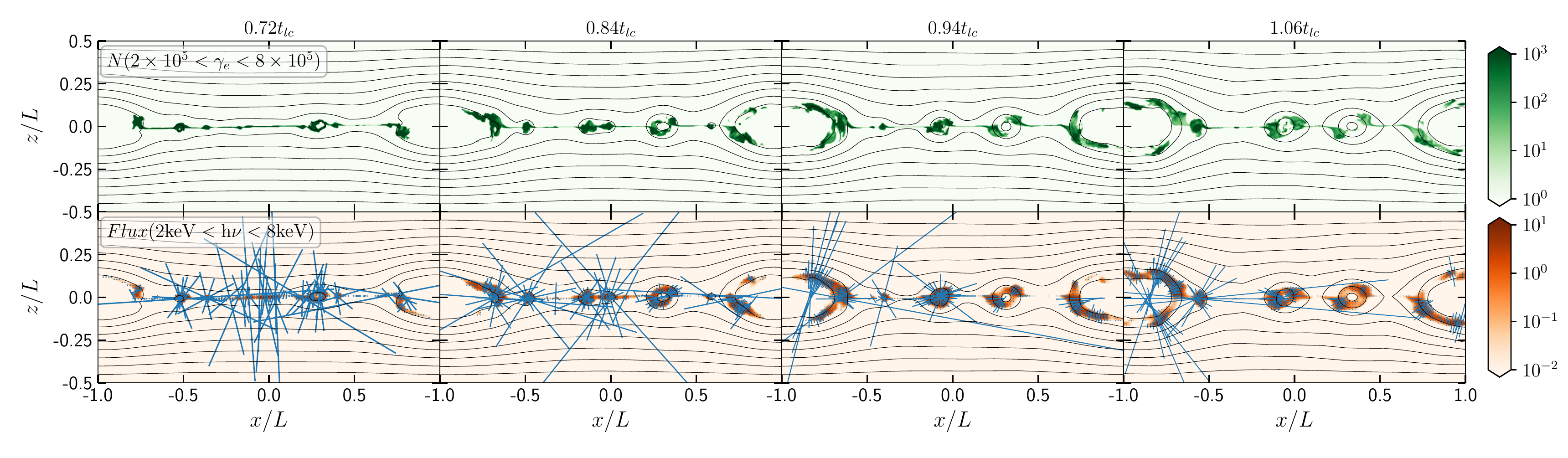}
\caption{Same as Figure \ref{fig:opticalsim} but for the X-ray band.}
\label{fig:xraysim}
\end{figure*}

\begin{figure*}
\centering
\includegraphics[width=0.99\textwidth]{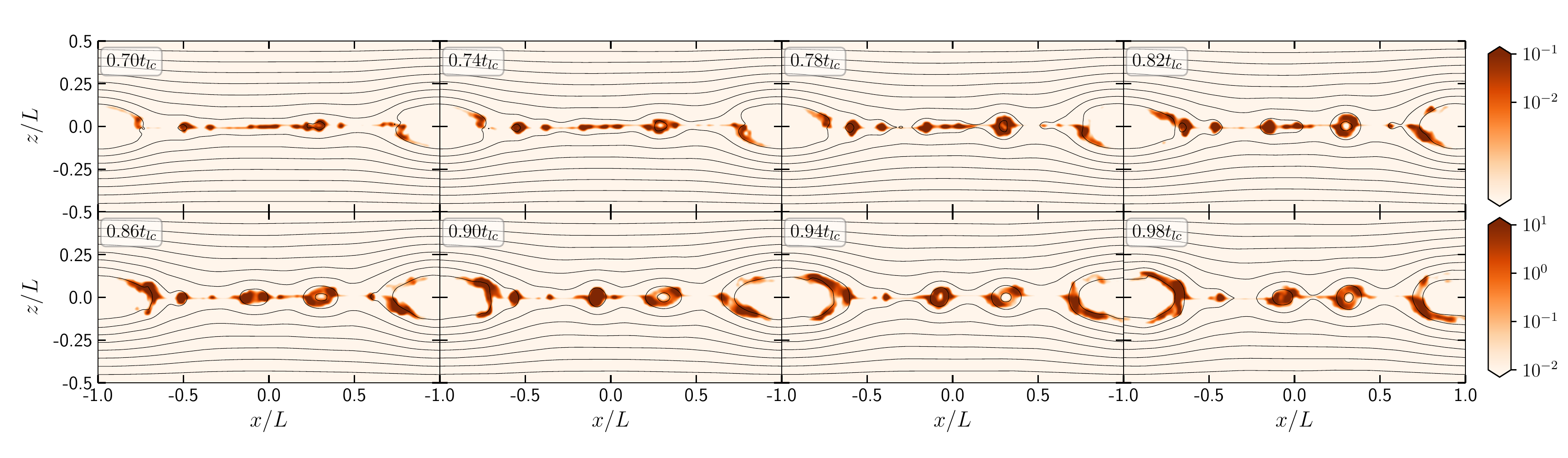}
\caption{Snapshots of the X-ray emission map during the first X-ray flare. The local polarized flux is not plotted so as to better show the size distribution of plasmoids.}
\label{fig:xrayzoom1}
\end{figure*}

\begin{figure*}
\centering
\includegraphics[width=0.99\textwidth]{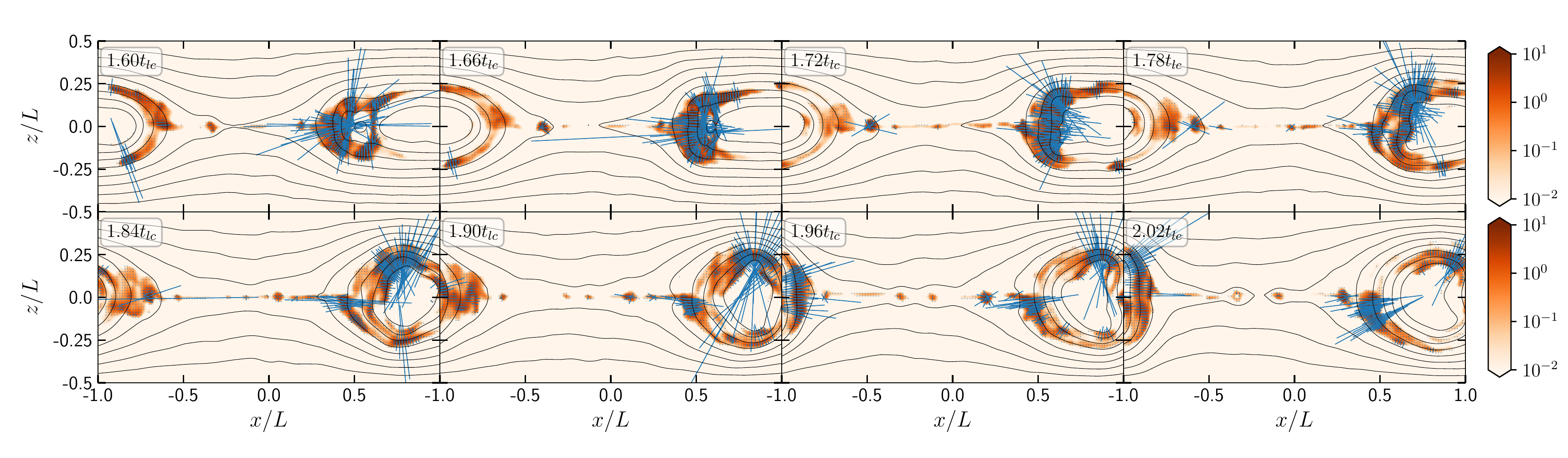}
\caption{Snapshots of the X-ray polarized emission map during the second X-ray flare.}
\label{fig:xrayzoom2}
\end{figure*}

\begin{figure*}
\centering
\includegraphics[width=0.99\textwidth]{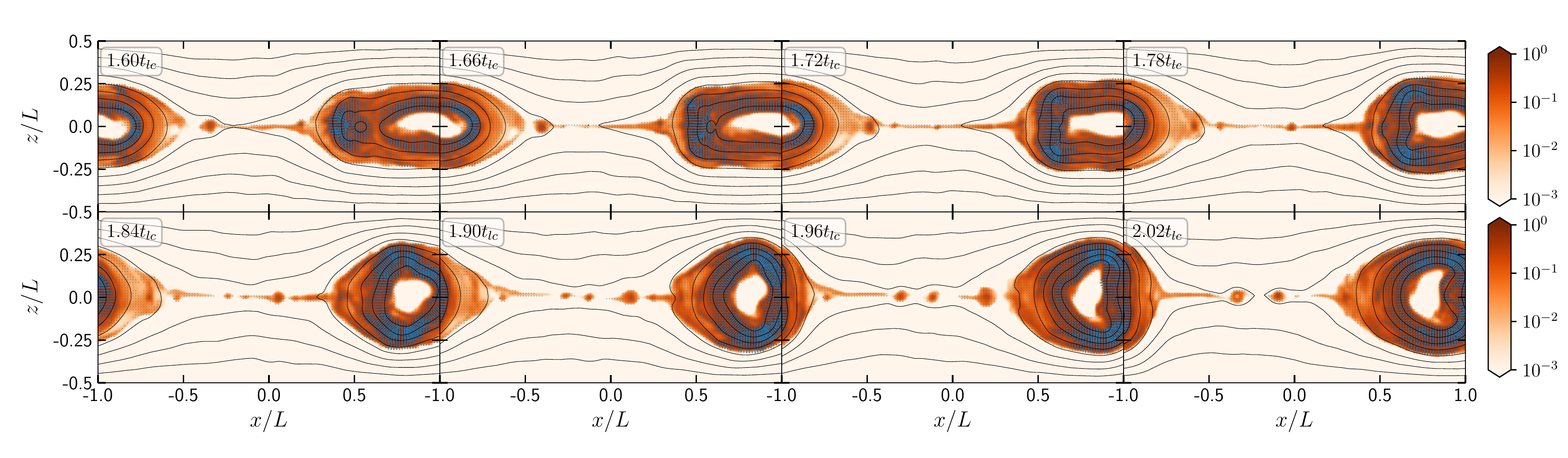}
\caption{Same as Figure \ref{fig:xrayzoom2} but for the optical band.}
\label{fig:opticalzoom2}
\end{figure*}


Figures \ref{fig:optical} and \ref{fig:xray} show the light curve, PD and PA for the optical and {\it IXPE} X-ray bands, respectively; Figures \ref{fig:opticalsim} and \ref{fig:xraysim} show snapshots of the particle spatial distribution and polarized emission map during the first X-ray flare (from $\sim0.7\tau_{lc}$ to  $\sim1.0\tau_{lc}$) for the two bands. The high magnetization factor and weak guide field lead to very efficient magnetic energy release by magnetic reconnection \citep{Guo2015,Li2017,Li2018a,Li2019a,Liu2020}. After the initial perturbation, the reconnection layer quickly produces a series of plasmoids, which are quasi-circular magnetic structures confining nonthermal particles (Figures \ref{fig:opticalsim} and \ref{fig:xraysim} upper panels). Due to the velocity difference, these plasmoids can merge into each other before colliding into the large plasmoid at the periodic boundary. Since the magnetic field lines around the plasmoids are all in the same direction, plasmoid mergers can lead to secondary magnetic reconnection at the contact region. This results in additional particle acceleration at the contact region of merging plasmoids, consistent with the previous findings \citep{Zhang2018,Zhang2020}. 

Most importantly, the electrons responsible for optical and X-ray emission suffer from distinct radiative cooling. One can quickly estimate the difference. The synchrotron critical frequency is proportional to $\gamma_e^2$, while the cooling time is proportional to $\gamma_e^{-1}$. Since the photon energy of the {\it IXPE} X-ray band is about 1000 times higher than the optical band, the electrons for the X-ray emission cools $\sim 30$ times faster than those for the optical. This is clearly shown in the light curves (Figures \ref{fig:optical} and \ref{fig:xray} upper panels): the optical light curve keeps rising until $t\sim 2 \tau_{lc}$, where it starts to cool; but the two X-ray flares both drop to half of the flare peak within $\sim 0.1\tau_{lc}$. Because of the strong cooling, the electrons responsible for X-ray emission cannot travel far from their acceleration sites. Figure \ref{fig:xraysim} (upper panel) shows that they locate mostly near the plasmoid center or at the contact region of plasmoid mergers. In contrast, those responsible for optical emission can fill up much larger regions of the reconnection layer (Figure \ref{fig:opticalsim}). Consequently, the X-ray emission represents the generation of many plasmoids and their mergers, which are very disordered. This leads to the highly variable X-ray light curve and similarly spiky PD and PA for the first flare, where the fast variability originates from small plasmoid mergers.
We suggest that the time scales of the fast variability patterns in both flux and polarization during the first X-ray flare are determined by the sizes of the plasmoids. Figure \ref{fig:xrayzoom1} shows snapshots of the X-ray emission maps in this period. We can clearly see that although the simulation box has periodic boundary in the outflow direction, the high-energy electrons responsible for X-ray emission have barely reached the boundary. The emission is mostly dominated by the central regions of plasmoids in the reconnection layer. By measuring the sizes of these X-ray shining regions, we find that they mostly fall in the region between $0.1L$ to $0.3L$, corresponding to $0.05 \tau_{lc}$ to $0.15 \tau_{lc}$. As shown in Figure \ref{fig:xray}, the time scales match very well with the fast variability patterns in both flux and polarization. We can also see in Figure \ref{fig:xraysim} that during the first X-ray flare, the local polarization vector distributions appear very stochastic without any systematic patterns. This explains the apparently random-walk X-ray PD and PA during the first flare (Figure \ref{fig:xray}).

The second X-ray flare, however, results from the merging of two large plasmoids. Figures \ref{fig:xrayzoom2} and \ref{fig:opticalzoom2} show the evolution of X-ray and optical polarized emission maps, respectively. As we can see in Figure \ref{fig:opticalzoom2}, at $t\sim 1.6 \tau_{lc}$, a large plasmoid starts to merge into the large plasmoid at the periodic boundary. This merger leads to strong particle acceleration at the contact region, as shown in the snapshot at $t\sim 1.66 \tau_{lc}$. However, just like the primary reconnection, the newly accelerated particles from this secondary reconnection site are not necessarily symmetric in the outflow direction, but there are more particles moving upwards than downwards. Since the magnetic field lines are quasi-circular in the post-merger plasmoid, the newly accelerated particles will stream along the magnetic field lines, so that the particles going upwards will stream clockwise, while those going downwards will go counter-clockwise. Due to the asymmetry, clockwise motion dominates the radiation signatures, which lights up the magnetic field structure along its trajectory, leading to a smooth PA rotation in the X-ray bands. These results are consistent with \citet{Zhang2018,Zhang2020}. One may expect that the large plasmoid merger should lead to a similar PA rotation in the optical band. However, since the electrons responsible for the optical emission cool much slower, they occupy a large region of the reconnection layer. As shown in Figure \ref{fig:opticalzoom2}, although the plasmoid merger accelerates a large amount of electrons, the local polarized flux is much less dominating compared to the X-ray band. In fact, the polarized flux is nearly symmetrically distributed in the quasi-circular plasmoid. The incoherent addition of the polarized emission thus cancels out the net polarization, making the optical PD drop to nearly zero. In this situation, a tiny excess in a specific direction of the polarized flux can make a huge impact on the net PA, which leads to the large but very narrow spike in the optical PA at $\sim 2.0\tau_{lc}$. However, since the optical PD is almost zero, this feature may not be observed. We want to emphasize that the large plasmoid mergers and resulting flare and PA swings are not always at the periodic boundary. As shown in \citet{Zhang2020}, the same can happen between two large plasmoids away from the boundary. Additionally, small plasmoid mergers in the relatively early stage of reconnection may lead to PA swings as well, while large plasmoid mergers do not necessarily lead to asymmetric outflows and PA swings. Nevertheless, our simulation has clearly shown that fast variability in flux and polarization as well as PA swings in the X-ray band can be characteristic signatures for magnetic reconnection in blazars.

The fast polarization variability in X-rays can be hard to resolve. Aforementioned, the temporal depolarization can diminish the detected polarization if the unresolved PA variation is large. To illustrate this effect, here we consider that one X-ray flare in our simulation is the period during which the relative flux is above one. Then the two flares at $\sim 0.75 \tau_{lc}$ and $\sim 1.7 \tau_{lc}$ are of similar duration ($\sim 0.5\tau_{lc}$). We consider three temporal resolutions, where the {\it IXPE} can resolve each flare by one, two, and three polarization points, respectively. The detected optical and X-ray PD and PA are shown in Figures \ref{fig:optical} and \ref{fig:xray} (middle and lower panels). For the first flare, the detected X-ray PD ($\sim 15\%$) is only $\sim 60\%$ of the optical ($\sim 25\%$) for any resolution. For the second flare, with one X-ray polarization point the PD ($\sim 9\%$) is still about $\sim 60\%$ of the optical ($\sim 15\%$), but with three points the X-ray PD is comparable to the optical and trace the intrinsic values well. Although three points have not yet resolved the PA rotation (Figure \ref{fig:xray} third panel from $\sim 1.5 \tau_{lc}$ to $\sim 2.0 \tau_{lc}$, and note the last green and orange PA data points are due to the $180^{\circ}$ PA ambiguity), during bright flares the {\it IXPE} may have better temporal resolution. The relatively smooth X-ray PA swing then can be a characteristic signature of reconnection.

\begin{figure}
\centering
\includegraphics[width=0.99\linewidth]{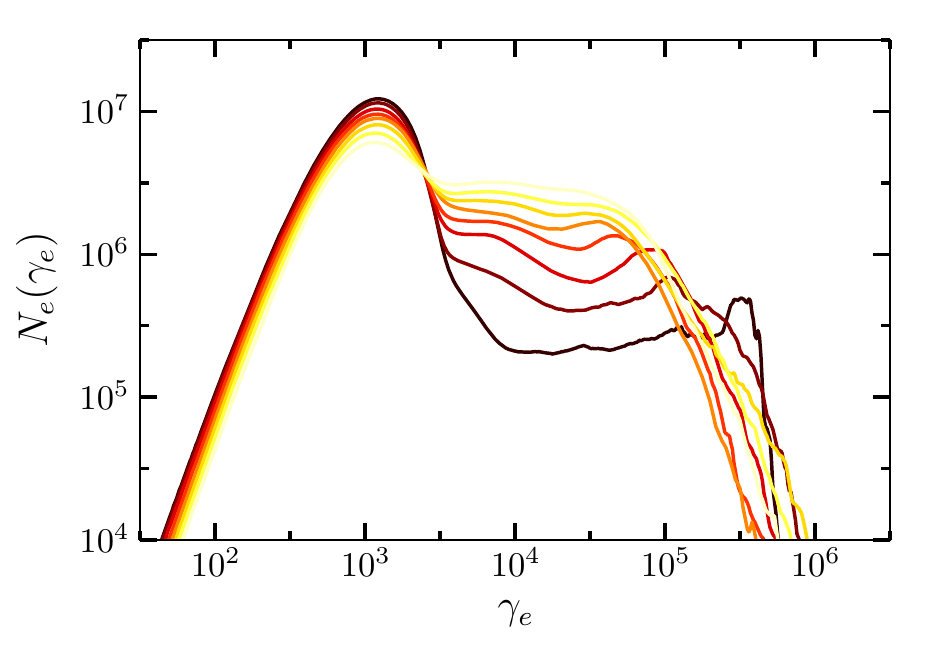}
\includegraphics[width=0.99\linewidth]{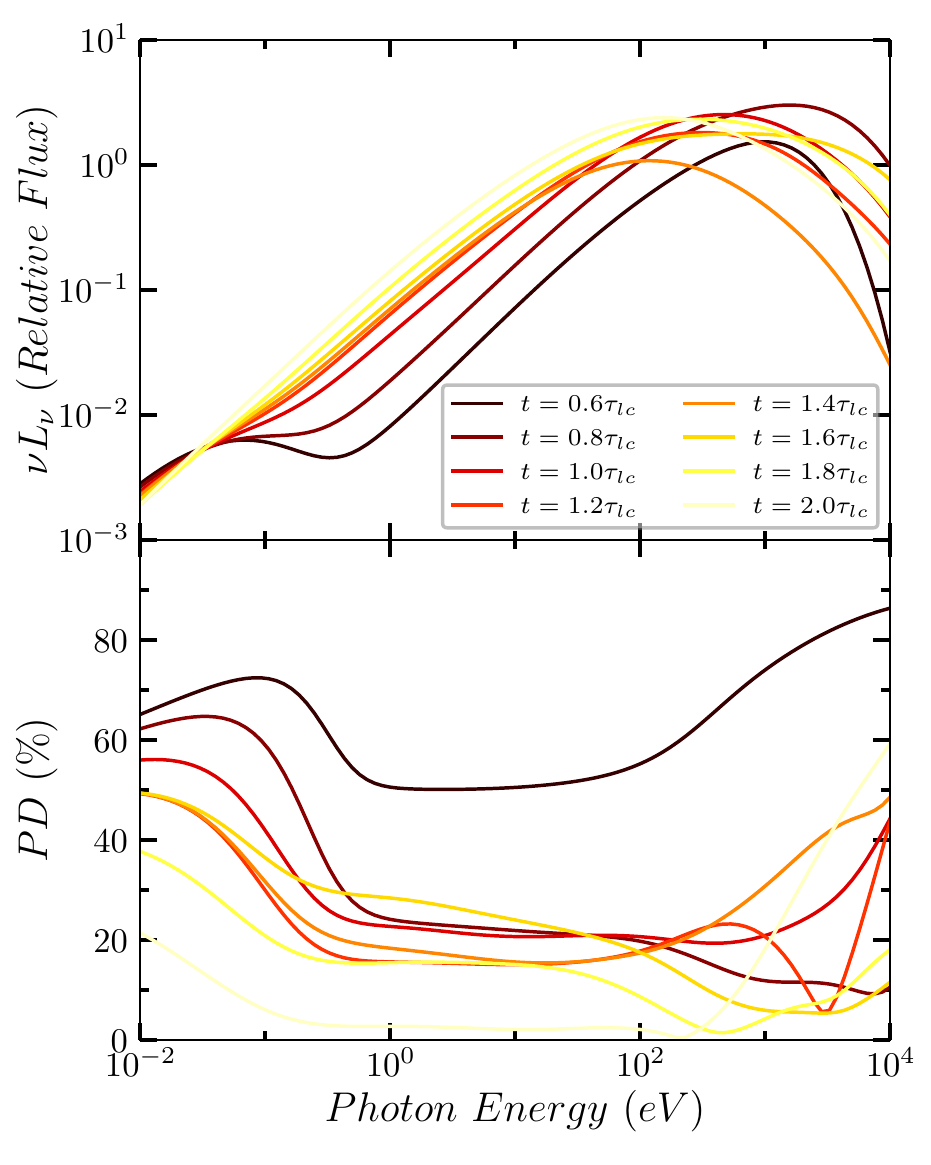}
\caption{From top to bottom are the snapshots of particle spectra, SEDs of the synchrotron component, and frequency-dependent PDs. SEDs are plotted in relative flux. The particle spectra are in the lab frame of the simulation, while the SEDs and PDs are in the observer's frame.}
\label{fig:spec}
\end{figure}

Figure \ref{fig:spec} shows the spectral properties of our simulation. We only plot the snapshots from approximately the starting of the first X-ray flare to the end of the second X-ray flare. We can clearly see that the reconnection quickly accelerates electrons into a power-law shape. Since the particle spectrum is very hard from the reconnection, the post-reconnection mean electron Lorentz factor is approximately at the cooling break $\gamma_c=10^5$, although this value can evolve in time due to the dynamical balance between acceleration and cooling. We note that the cooling break here comes from our normalization so that it fits with typical HBL observations, similar to \citet{Zhang2018,Zhang2020}. This is through the scale-up of the radiative cooling, and we calculate that the so-called synchrotron burn-off limit is at $\gamma_b\sim 10^6$ in our simulation \citep{Uzdensky2011}, which is evident from particle spectrum that cuts off at slightly lower than $\gamma=10^6$. We can see that the synchrotron spectrum is generally harder and extends to higher energies during flare peaks ($0.8$ to $1.0 \tau_{lc}$ and $1.6$ to $1.8 \tau_{lc}$ in the middle panel of Figure \ref{fig:spec}) than low-states ($1.2$ to $1.4 \tau_{lc}$). Additionally, we can see that the PD variation in optical to UV bands, which is before the cooling break, is much smaller than that in the X-ray bands, which is beyond the cooling break. As mentioned previously, this is due to the radiative cooling, because electrons responsible for X-ray emission only occupy the very active regions in the reconnection layer. All these spectral properties are consistent with our previous findings with flat spectrum radio quasars \citep{Zhang2018,Zhang2020}.

\section{Implication for Observations \label{sec:discussion}}

To summarize, for the first time we use combined PIC and polarized radiation transfer simulation to study time-dependent optical and X-ray polarization signatures from magnetic reconnection in HBLs. Our study is unique in that we predict polarization signatures from first principles and take into account the often overlooked temporal depolarization effect, which is essential for X-ray polarimetry. We find that the reconnection-driven X-ray light curves show much stronger variability than the optical band, owing to its much faster radiative cooling. Similarly, the intrinsic evolution of the X-ray PD and PA also shows significant micro-variability. During X-ray flares, the average intrinsic optical and X-ray PDs are comparable. However, depending on the time resolution of the {\it IXPE} data, the detected X-ray PD can be as low as $\sim 60\%$ of the optical PD in the case of low resolution; in the case of high resolution, the optical and X-ray PDs are comparable, and we may observe X-ray PA swings. These polarization signatures are characteristic of reconnection in HBLs.

Under the magnetic reconnection scenario, the blazar polarization variability of a specific observational band depends on its position in the SED. This results from the combined effect of radiative cooling and highly dynamical magnetic field evolution \citep[also see][]{Zhang2020}. For observational bands below the synchrotron peak, the cooling of electrons is relatively slow, thus they can fill up plasmoids in the reconnection region. The resulting PD and PA are usually small fluctuations around some mean values, unless there are large plasmoid mergers, where both the PD and PA may experience large changes. On the other hand, the electrons responsible for observational bands near or beyond the synchrotron peak quickly cool after they are accelerated. Therefore, they mostly exist near sites with strong acceleration, such as plasmoid mergers. The resulting polarization signatures thus appear highly variable. Consequently, strongly variable polarization near the synchrotron peak of a blazar can be a signature of the magnetic reconnection. An interesting inference is that optical polarization observations may expect a ``blazar polarization sequence'', where the optical polarization is less variable for higher-frequency blazars. This feature can be tested statistically to identify whether reconnection is the dominant driver of blazar flares.

Another interesting implication is that the temporal depolarization may lower the blazar optical polarization as well. Observations have shown that the optical PD appears lower during PA swings \citep[according to][Figure 9, the PD during swings is about $\sim 70\%$ of the average]{Blinov2016}. However, in many flaring events observations can only grab a few data points during the PA swing. As illustrated in Figure \ref{fig:tempdepol}, the temporal depolarization by itself can lead to this polarization reduction. Therefore, we need high cadence optical polarization monitoring to test whether the polarization reduction during swings is due to a more disordered magnetic field or merely temporal depolarization.

We want to mention a few caveats in our simulations. Our simulations start with a preexisting current sheet, which may only represent one emission region for one flaring epoch. In reality, Mrk~421 and 501 are fairly bright in X-ray bands even during low states \citep{Fraija2017}. This indicates additional emission regions in the jet that are not simulated in our work. Such extra emission regions may contribute to a ``quiescent'' flux level, so that the flare amplitude shown in our simulations may be lower. Additionally, the quiescent flux may also contaminate the observed polarization signatures. Optical polarimetry often detects $\lesssim 10\%$ PD for both flaring and quiescent states \citep{Hovatta2016,Fraija2017}. In our simulations, however, we see $\sim 20\%$ optical PD. This is because the 2D PIC simulation cannot fully capture turbulence in the reconnection region. \citet{Li2019} and \citet{Guo2020} have shown that turbulence is very strong in 3D reconnection simulations. This can lead to further reduction in both optical and X-ray polarization. Given a flaring event with $\sim 10\%$ optical polarization, the detected X-ray polarization considering the temporal depolarization may be as low as $\sim 5\%$. Consider the flux of Mrk 421 at $10^{-10}~\rm{erg\,cm^{-2}s^{-1}}$ in the $2-8~\rm{keV}$ band with a photon index of 2.5, the {\it IXPE} will take about $100~\rm{ks}$ to obtain a minimal detectable polarization of $\sim 4\%$\footnote{The numbers are based on the {\it IXPE} WebPIMMS tool at \url{https://ixpe.msfc.nasa.gov/cgi-aft/w3pimms/w3pimms.pl}}. Therefore, if the emission of Mrk~421 and 501 are driven by magnetic reconnection, the {\it IXPE} may report very low X-ray PD or even non-detection when the sources are in the quiescent state. During bright flares, however, {\it IXPE} can have much better time resolution. Assuming that the flux of Mrk 421 can reach $10^{-9}~\rm{erg\,cm^{-2}s^{-1}}$, then {\it IXPE} can obtain a minimal detectable polarization of $\sim 4\%$ within $10~\rm{ks}$, which may be able to fully resolve the fast variability in polarization and X-ray PA swing.

Therefore, we suggest that the {\it IXPE} has the potential to distinguish the shock, turbulence, and magnetic reconnection scenarios via X-ray polarimetry. The shock scenario predicts very stable X-ray polarization with much higher PD than the optical band, contrary to the reconnection scenario \citep{Tavecchio2018,Tavecchio2020}. While the frequency-dependent polarization has not been well studied for the turbulence scenario, generally it is unlikely to produce rather smooth PA swings \citep{Marscher2014,Kiehlmann2017,Marscher2017}, which come from plasmoid mergers and may be resolved during bright X-ray flares. On the other hand, reconnection predicts very low and stochastic polarization signatures when the source is not very active, but strongly variable flux and polarization, as well as potential X-ray PA swing, during flaring states. These unique signatures can be examined by future {\it IXPE} polarimetry.

\acknowledgments{
We thank the anonymous referee for very helpful and constructive reviews. H. Z. and D. G. acknowledge support from the NASA ATP NNX17AG21G, the NSF AST-1910451, the NSF AST-1816136 grants and by Fermi Cycle 12 Guest Investigator Program \#121077. The work by H. Z. and X. L. is funded by the National Science Foundation (NSF) grant PHY-1902867 and Department of Energy (DOE) DE-SC0020219 through the NSF/DOE Partnership in Basic Plasma Science and Engineering.  F.
G. acknowledges support in part from Astrophysics Theory Program, and DOE support through the LDRD program at LANL.}


\begin{thebibliography}{99}
\bibitem[Ahnen et al.(2018)]{Ahnen2018} Ahnen, M.~L., Ansoldi, S., Antonelli, L.~A., et al.\ 2018, \aap, 620, A181. doi:10.1051/0004-6361/201833704

\bibitem[Albert et al.(2007a)]{Albert2007a} Albert, J., Aliu, E., Anderhub, H., et al.\ 2007a, \apj, 663, 125. doi:10.1086/518221
\bibitem[Albert et al.(2007b)]{Albert2007b} Albert, J., Aliu, E., Anderhub, H., et al.\ 2007b, \apj, 669, 862. doi:10.1086/521382
\bibitem[Aleksi{\'c} et al.(2015)]{Aleksic2015} Aleksi{\'c}, J., Ansoldi, S., Antonelli, L.~A., et al.\ 2015, \aap, 578, A22. doi:10.1051/0004-6361/201424811
\bibitem[B{\l}a{\.z}ejowski et al.(2005)]{Blazejowski2005} B{\l}a{\.z}ejowski, M., Blaylock, G., Bond, I.~H., et al.\ 2005, \apj, 630, 130. doi:10.1086/431925


\bibitem[Blinov et al.(2016)]{Blinov2016} Blinov, D., Pavlidou, V., Papadakis, I.~E., et al.\ 2016, \mnras, 457, 2252. doi:10.1093/mnras/stw158

\bibitem[Bowers et al.(2008)]{Bowers2008} Bowers, K.~J., Albright, B.~J., Yin, L., et al.\ 2008, Physics of Plasmas, 15, 055703. doi:10.1063/1.2840133
\bibitem[Chen et al.(2011)]{Chen2011} Chen, X., Fossati, G., Liang, E.~P., et al.\ 2011, \mnras, 416, 2368. doi:10.1111/j.1365-2966.2011.19215.x
\bibitem[Christie et al.(2020)]{Christie2020} Christie, I.~M., Petropoulou, M., Sironi, L., et al.\ 2020, \mnras, 492, 549. doi:10.1093/mnras/stz3265

\bibitem[Finke(2013)]{Finke2013} Finke, J.~D.\ 2013, \apj, 763, 134. doi:10.1088/0004-637X/763/2/134

\bibitem[Fraija et al.(2017)]{Fraija2017} Fraija, N., Ben{\'\i}tez, E., Hiriart, D., et al.\ 2017, \apjs, 232, 7. doi:10.3847/1538-4365/aa82cc

\bibitem[Giannios et al.(2009)]{Giannios2009} Giannios, D., Uzdensky, D.~A., \& Begelman, M.~C.\ 2009, \mnras, 395, L29. doi:10.1111/j.1745-3933.2009.00635.x
\bibitem[Giannios \& Uzdensky(2019)]{Giannios2019} Giannios, D. \& Uzdensky, D.~A.\ 2019, \mnras, 484, 1378. doi:10.1093/mnras/stz082

\bibitem[Guo et al.(2014)]{Guo2014} Guo, F., Li, H., Daughton, W., et al.\ 2014, \prl, 113, 155005. doi:10.1103/PhysRevLett.113.155005

\bibitem[Guo et al.(2015)]{Guo2015} Guo, F., Liu, Y.-H., Daughton, W., et al.\ 2015, \apj, 806, 167. doi:10.1088/0004-637X/806/2/167


\bibitem[Guo et al.(2016)]{Guo2016} Guo, F., Li, X., Li, H., et al.\ 2016, \apjl, 818, L9. doi:10.3847/2041-8205/818/1/L9

\bibitem[Guo et al.(2020a)]{Guo2020a} Guo, F., Liu, Y.-H., Li, X., et al.\ 2020, Physics of Plasmas, 27, 080501. doi:10.1063/5.0012094


\bibitem[Guo et al.(2020b)]{Guo2020} Guo, F., Li, X., Daughton, W., et al.\ 2020, arXiv:2008.02743

\bibitem[Hosking \& Sironi(2020)]{Hosking2020} Hosking, D.~N. \& Sironi, L.\ 2020, \apjl, 900, L23. doi:10.3847/2041-8213/abafa6
\bibitem[Hovatta et al.(2016)]{Hovatta2016} Hovatta, T., Lindfors, E., Blinov, D., et al.\ 2016, \aap, 596, A78. doi:10.1051/0004-6361/201628974

\bibitem[Kiehlmann et al.(2017)]{Kiehlmann2017} Kiehlmann, S., Blinov, D., Pearson, T.~J., et al.\ 2017, \mnras, 472, 3589. doi:10.1093/mnras/stx2167

\bibitem[Kilian et al.(2020)]{Kilian2020} Kilian, P., Li, X., Guo, F., et al.\ 2020, \apj, 899, 151. doi:10.3847/1538-4357/aba1e9

\bibitem[Li et al.(2017)]{Li2017} Li, X., Guo, F., Li, H., et al.\ 2017, \apj, 843, 21. doi:10.3847/1538-4357/aa745e

\bibitem[Li et al.(2018)]{Li2018a} Li, X., Guo, F., Li, H., et al.\ 2018a, \apj, 855, 80. doi:10.3847/1538-4357/aaacd5


\bibitem[Li et al.(2018)]{Li2018} Li, X., Guo, F., Li, H., et al.\ 2018b, \apj, 866, 4. doi:10.3847/1538-4357/aae07b

\bibitem[Li et al.(2019)]{Li2019a} Li, X., Guo, F., \& Li, H.\ 2019a, \apj, 879, 5. doi:10.3847/1538-4357/ab223b


\bibitem[Li et al.(2019)]{Li2019} Li, X., Guo, F., Li, H., et al.\ 2019b, \apj, 884, 118. doi:10.3847/1538-4357/ab4268

\bibitem[Liu et al.(2020)]{Liu2020} Liu, Y.-H., Lin, S.-C., Hesse, M., et al.\ 2020, \apjl, 892, L13. doi:10.3847/2041-8213/ab7d3f

\bibitem[Marscher(2014)]{Marscher2014} Marscher, A.~P.\ 2014, \apj, 780, 87. doi:10.1088/0004-637X/780/1/87

\bibitem[Marscher et al.(2017)]{Marscher2017} Marscher, A., Jorstad, S., \& Williamson, K.\ 2017, Galaxies, 5, 63. doi:10.3390/galaxies5040063

\bibitem[Padovani \& Giommi(1995)]{Padovani1995} Padovani, P. \& Giommi, P.\ 1995, \apj, 444, 567. doi:10.1086/175631

\bibitem[Sironi \& Spitkovsky(2014)]{Sironi2014} Sironi, L. \& Spitkovsky, A.\ 2014, \apjl, 783, L21. doi:10.1088/2041-8205/783/1/L21

\bibitem[Sironi et al.(2016)]{Sironi2016} Sironi, L., Giannios, D., \& Petropoulou, M.\ 2016, \mnras, 462, 48. doi:10.1093/mnras/stw1620


\bibitem[Tavecchio et al.(2018)]{Tavecchio2018} Tavecchio, F., Landoni, M., Sironi, L., et al.\ 2018, \mnras, 480, 2872. doi:10.1093/mnras/sty1491

\bibitem[Tavecchio et al.(2020)]{Tavecchio2020} Tavecchio, F., Landoni, M., Sironi, L., et al.\ 2020, \mnras, 498, 599. doi:10.1093/mnras/staa2457
\bibitem[Uzdensky et al.(2011)]{Uzdensky2011} Uzdensky, D.~A., Cerutti, B., \& Begelman, M.~C.\ 2011, \apjl, 737, L40. doi:10.1088/2041-8205/737/2/L40

\bibitem[Werner et al.(2016)]{Werner2016} Werner, G.~R., Uzdensky, D.~A., Cerutti, B., et al.\ 2016, \apjl, 816, L8. doi:10.3847/2041-8205/816/1/L8


\bibitem[Zhang et al.(2014)]{Zhang2014} Zhang, H., Chen, X., \& B{\"o}ttcher, M.\ 2014, \apj, 789, 66. doi:10.1088/0004-637X/789/1/66
\bibitem[Zhang et al.(2015)]{Zhang2015} Zhang, H., Chen, X., B{\"o}ttcher, M., et al.\ 2015, \apj, 804, 58. doi:10.1088/0004-637X/804/1/58
\bibitem[Zhang et al.(2017)]{Zhang2017} Zhang, H., Li, H., Guo, F., et al.\ 2017, \apj, 835, 125. doi:10.3847/1538-4357/835/2/125

\bibitem[Zhang et al.(2018)]{Zhang2018} Zhang, H., Li, X., Guo, F., et al.\ 2018, \apjl, 862, L25. doi:10.3847/2041-8213/aad54f
\bibitem[Zhang et al.(2020)]{Zhang2020} Zhang, H., Li, X., Giannios, D., et al.\ 2020, \apj, 901, 149. doi:10.3847/1538-4357/abb1b0
\bibitem[Zhang \& Giannios(2021)]{Zhang2021} Zhang, H. \& Giannios, D.\ 2021, \mnras, 502, 1145. doi:10.1093/mnras/stab008


\end{thebibliography}



\end{document}